\documentclass[]{ecai} 

\usepackage{latexsym}
\usepackage{amssymb}
\usepackage{amsmath}
\usepackage{amsthm}
\usepackage{booktabs}
\usepackage{enumitem}
\usepackage{graphicx}
\usepackage{color}
\usepackage{graphicx} 
\usepackage{algorithm}
\usepackage{algorithmic}

\usepackage[utf8]{inputenc} 
\usepackage[T1]{fontenc}    
\usepackage{url}            
\usepackage{booktabs}       
\usepackage{amsfonts}       
\usepackage{nicefrac}       
\usepackage{microtype}      
\usepackage{xcolor}         
\usepackage{multirow}%
\usepackage{tabularray} 
\usepackage{amsfonts}
\usepackage{amsmath}
\usepackage{fancyhdr}
\usepackage{graphicx}
\usepackage{caption}
\usepackage{subcaption}
\usepackage{caption}
\usepackage{listings}
\usepackage{tabularx}

\usepackage{newfloat}
\usepackage{listings}

\newcommand{\BibTeX}{B\kern-.05em{\sc i\kern-.025em b}\kern-.08em\TeX}

\begin{document}


\begin{frontmatter}



\title{TLOB: A Novel Transformer Model with Dual Attention for Price Trend Prediction with Limit Order Book Data}


\author[A]{\fnms{Leonardo}~\snm{Berti}\thanks{Corresponding Author. Email: berti.1883894@studenti.uniroma1.it}}
\author[B]{\fnms{Gjergji}~\snm{Kasneci}}

\address[A]{Sapienza University of Rome}
\address[B]{Technical University of Munich}

\paperid{7943}

\begin{abstract}
Price Trend Prediction (PTP) based on Limit Order Book (LOB) data is a fundamental challenge in financial markets. Despite advances in deep learning, existing models fail to generalize across different market conditions and assets. Surprisingly, by adapting a simple MLP-based architecture to LOB, we show that we surpass SoTA performance; thus, challenging the necessity of complex architectures. Unlike past work that shows robustness issues, we propose TLOB, a transformer-based model that uses a dual attention mechanism to capture spatial and temporal dependencies in LOB data. This allows it to adaptively focus on the market microstructure, making it particularly effective for longer-horizon predictions and volatile market conditions.
We also introduce a new labeling method that improves on previous ones, removing the horizon bias.
We evaluate TLOB's effectiveness across four horizons, using the established FI-2010 benchmark, which exceeds the state-of-the-art by an average of 3.7 F1-score. Additionally, TLOB shows average improvements on Tesla and Intel with a 1.3 and 7.7 increase in F1-score, respectively. Finally, we tested TLOB on a recent Bitcoin dataset, and TLOB outperforms the SoTA performance by an average of 1.1 in F1-score.
Additionally, we empirically show how stock price predictability has declined over time, -6.68 in F1-score, highlighting the growing market efficiency. 
Predictability must be considered in relation to transaction costs, so we experimented with defining trends using an average spread, reflecting the primary transaction cost. The resulting performance deterioration underscores the complexity of translating trend classification into profitable trading strategies.
We argue that our work provides new insights into the evolving landscape of stock price trend prediction and sets a strong foundation for future advancements in financial AI. We commit to releasing the code publicly. 
\end{abstract}

\end{frontmatter}
\section{Introduction}
Over the past few decades, the global financial landscape has undergone a profound transformation, transitioning from manual trading operations to sophisticated electronic platforms. This evolution has been so significant that by 2020, electronic trading accounted for over 99\% of equity shares traded in the United States, a stark contrast to just 15\% in 2000 \cite{kissell2020algorithmic}. At the heart of this revolution lies the electronic Limit Order Book (LOB), a dynamic data structure that has become the cornerstone of modern financial markets.
In today's competitive financial world, the majority of the markets utilize electronic LOBs to record trades. 
The continuous inflow of limit orders, organized by price levels, creates a dynamic structure that evolves over time, reflecting the real-time balance of supply and demand. However, this multidimensional structure, which spans price levels and volumes, presents complex challenges for understanding market behavior, forecasting stock price trends, and simulating realistic market conditions.
The non-stationary nature of the markets, characterized by their stochastic behavior, makes modeling security price movements challenging. Traditional statistical methods fail to capture these complexities, especially when attempting to predict short-term price trends. However, recent advancements in deep learning have opened new avenues for tackling these challenges, offering the ability to model the non-linear relationships and temporal dependencies inherent in LOB data.

Price Trend Prediction (PTP)\footnote{In the literature, it is also referred to as mid-price movement prediction.} remains one of the most challenging and economically significant problems in financial markets, attracting significant attention from academic researchers and industry practitioners. One prominent application of PTP, particularly utilizing Limit Order Book (LOB) data, lies within high-frequency trading, where algorithms attempt to capitalize on short-term price movements. Predicting future market movements is a highly challenging task due to the complexity, non-stationarity, and volatility of financial markets. However, with the growing availability of Limit Order Book (LOB) data and advancements in deep learning, new opportunities have emerged to improve the accuracy of these predictions. This paper explores the application of deep learning models to PTP using Limit Order Book (LOB) data, which provides the most granular and complete information on trades.
Financial markets do not exist in a vacuum; they are continuously shaped by the actions and expectations of countless participants who, according to the Efficient Market Hypothesis (EMH), collectively incorporate all available information into asset prices. When models discover a predictive pattern and traders act on it, the anomaly is quickly competed away, causing a paradox: successful signals sow the seeds of their own demise. Over time, greater liquidity, advanced trading technologies, and the proliferation of algorithmic strategies intensify this effect, i.e., any exploitable signal becomes visible in execution data and erodes more rapidly. Consequently, the apparent decline in forecast accuracy in our findings aligns with EMH principles: as soon as new patterns are detected and exploited, the relentless engine of arbitrage drives markets back toward efficiency. This interplay underlines why forecasting often becomes more difficult the farther we move from idealized, less liquid markets (like FI-2010) toward active, high-efficiency markets (like NASDAQ), thereby illustrating a core tension between the pursuit of alpha and the self-correcting nature of competitive markets. 
Traditional forecasting approaches relied on technical analysis and statistical methods, but recent years have seen a shift toward more sophisticated deep learning methods.
A lot of different types of deep learning architectures have been utilized to tackle the PTP tasks. Recurrent Neural Networks (RNNs) based on Long-Short Term Memory (LSTM) layers \cite{tsantekidis2017forecasting}, Convolutional Neural Networks (CNNs) \cite{tsantekidis2017using, tsantekidis2020using, zhang2019deeplob}, Temporal Attention-Augmented Bilinear architecture (TABL) \cite{tran2018temporal}, and many others \cite{kisiel2022axial, zhang2021multihorizonforecastinglimitorder, passalis2019deep}.
Recent work \cite{prata2024lob} has highlighted the limitations of existing models, particularly their lack of robustness and generalizability when applied to diverse market conditions and more efficient stocks.
In this paper, we address these limitations by proposing TLOB, a transformer-based approach that outperforms all the existing models on both benchmark and real-world datasets, paving the way for more reliable PTP applications. We also introduce an MLP-based model to show that a simple architecture, based on fully connected layers and GeLU activation function, can outperform all the SoTA models.
We list our contributions:
\begin{enumerate}
    \item \textbf{Novel Architecture Proposals}: We introduce two new deep learning models that surpass state-of-the-art performances:
    \begin{itemize}
        \item \textbf{MLPLOB}: A simple yet effective MLP-based model inspired by recent advances in the deep learning literature.
        \item \textbf{TLOB}: A transformer-based approach that leverages dual attention mechanisms for both temporal and spatial relationships in LOB data.
    \end{itemize}
    \item \textbf{Comprehensive Evaluation}: We conduct extensive experiments on the benchmark FI-2010 dataset, a NASDAQ dataset composed of Tesla and Intel stocks, and a BTC dataset, with several baselines, providing insights into model performance across different market assets, conditions, and horizons. We also perform an ablation study investigating the design choices of TLOB. 
    \item \textbf{New Labeling Methods}: We introduce a new labeling method that improves on previous ones, removing the horizon bias.
    \item \textbf{Historical Comparison}: We examine whether stock price prediction has become more difficult over time by comparing model performance on historical data from different periods. 
    \item \textbf{Alternative Threshold Definition}: We propose and evaluate a novel approach to defining trend classification thresholds based on average spread, directly incorporating the primary transaction cost into the prediction framework. 

\end{enumerate}

\section{Background}
In the contemporary, highly competitive financial landscape, the predominant mechanism for recording and managing market transactions is the electronic Limit Order Book (LOB). Within a limit order book market, traders can submit orders to buy or sell a specified quantity of an asset at a predetermined price. Three primary order types are prevalent in such markets: (1) \textbf{Market orders}, which are executed immediately at the best available price with a predetermined quantity; (2) \textbf{Limit orders}, allows traders to decide the maximum (in the case of a buy) or the minimum (in the case of a sell) price at which they want to complete the transaction. A quantity is always associated  with the specified price; and (3) \textbf{Cancel orders} (alternatively referred to as deletions), which serve to remove an active limit order.

The LOB is a data structure that maintains and matches active limit orders and market orders in accordance with a predefined set of rules. This structure is transparently accessible to all market participants and is subject to continuous updates with each event, including order placement, modification, cancellation, and execution. The most widely adopted mechanism for order matching is the Continuous Double Auction (CDA) \cite{bouchaud2018trades}. Under the CDA framework, orders are executed whenever the best bid (the highest price a buyer is willing to offer) and the best ask (the lowest price a seller is willing to accept) overlap. This mechanism facilitates continuous and competitive trading among market participants. The price of a security is commonly defined as the mid-price, calculated as the average of the best ask and best bid prices, with the difference between these prices representing the bid-ask spread.

Given that limit orders are organized into distinct depth levels, each comprising bid price, bid size, ask price, and ask size, based on their respective prices, the temporal evolution of a LOB constitutes a complex, multidimensional temporal problem. Research on LOB data can be broadly categorized into four primary types: empirical analyses of LOB dynamics \cite{cont2001empirical, bouchaud2002statistical}, price and volatility forecasting \cite{zhang2019deeplob, sirignano2019deep}, stochastic modeling of LOB dynamics \cite{cont2011statistical, gould2013limit}, and LOB market simulation \cite{byrd2020abides, coletta2021towards, li2020generating}.

\section{Related Work}
The task of accurately modeling the complex data patterns and large volumes linked to Limit Order Books (LOBs) has driven the advancement of deep learning algorithms.
In this section, we will summarize the State-of-The-Art (SoTA) deep learning models in the Price Trend Prediction (PTP) task. 
Tsantekidis et al. initially employed Recurrent Neural Networks (RNNs) based on Long-Short Term Memory (LSTM) layers \cite{tsantekidis2017using} and subsequently introduced a CNN-based model (CNN) \cite{tsantekidis2017forecasting}. Later, they proposed CNN-LSTM \cite{tsantekidis2020using}, which combines CNN feature extraction with LSTM classification.
Tran et al. developed the Temporal Attention-Augmented Bilinear Layer (TABL) \cite{tran2018temporal} for multivariate time series, capturing feature and temporal dependencies via bilinear transformations. This was extended to BINCTABL \cite{tran2021data}, incorporating a bilinear normalization layer to handle non-stationarity and magnitude disparities.
Passalis et al. introduced DAIN (Deep Adaptive Input Normalization) \cite{passalis2019deep}, a three-step adaptive normalization layer (shifting, scaling, gating), which was integrated into MLPs, CNNs, and RNNs.
Zhang et al. presented DEEPLOB \cite{zhang2019deeplob}, comprising convolutional layers, an Inception Module for feature extraction, and an LSTM for temporal dependencies. They later enhanced this with an attention mechanism in DEEPLOBATT \cite{zhang2021multihorizonforecastinglimitorder} for multi-horizon forecasting, using an encoder and attention-weighted hidden states.
Kiesel et al. introduced Axial-LOB \cite{kisiel2022axial}, utilizing axial attention to factorize 2D attention into separate 1D modules for feature and time axes.
For a comprehensive review and benchmark of those models we refer the reader to \cite{prata2024lob}.

\section{Task Definition}
\label{sec:task}

We represent the evolution of a LOB as a time series $\mathbb{L}$, where each $\mathbb{L}(t) \in \mathbb{R}^{4L}$ is called a LOB record, for $t=1,\ldots,N$, with $N$ being the number of LOB observations and $L$ the number of levels. In particular,
\begin{equation}
\mathbb{L}(t) = \bigl(P^{ask}(t),\, V^{ask}(t),\, P^{bid}(t),\, V^{bid}(t)\bigr),
\end{equation}
where $P^{ask}(t)$ and $P^{bid}(t)\in \mathbb{R}^L$ are the prices at levels $1$ through $L$, and $V^{ask}(t)$ and $V^{bid}(t)\in \mathbb{R}^L$ are the corresponding volumes.

\textbf{Trend Definition}
\label{sec:data_labelling}
We employ a ternary classification system for price trends: 
\texttt{U} (``upward'') denotes an increasing price trend, 
\texttt{D} (``downward'') indicates a decreasing trend, 
and \texttt{S} (``stable'') represents price movements with only minor variations.

In equity markets, mid-prices are generally considered the most reliable single-value indicator of actual asset prices. However, owing to inherent market fluctuations and exogenous shocks, mid-prices can exhibit considerable volatility. Consequently, labeling consecutive mid-prices $\bigl(p_t, p_{t+1}\bigr)$ often results in noisy labels.
\begin{figure}
    \centering
    \includegraphics[width=0.8\linewidth]{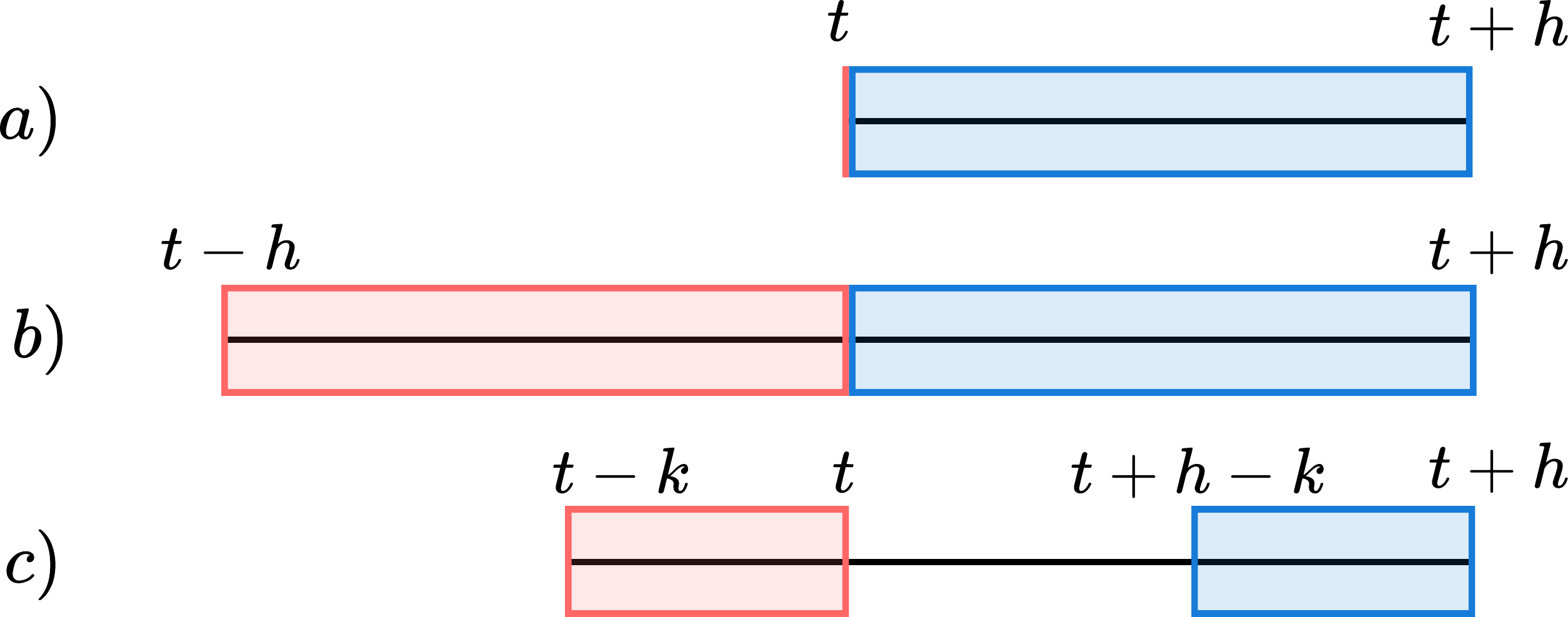}
    \vspace{0.3cm}
    \caption{Comparison of three labeling methods. $t$ is the current timestamp, $k$ is the smoothing window length, and $h$ is the prediction horizon. In our proposed method (c), $k$ and $h$ are defined independently, providing a more flexible and unbiased approach.}
    \label{fig:labeling}
\end{figure}
\\
\\
\\
\\
To mitigate this, many labeling strategies employ smoother mid-price functions, averaging prices over a chosen ``window length'' to reduce short-term noise and better reflect persistent directional moves. An example of this approach appears in \cite{urn:nbn:fi:csc-kata20170601153214969115}, detailed in Section~\ref{sec:fi-2010}.
However, as shown by Zhang et al.~\cite{zhang2019deeplob} (Fig.~2), smoothing only the future prices can lead to instability in trading signals. This instability often causes redundant trading actions and higher transaction costs. To address this, Tsantekidis et al.~\cite{tsantekidis2017forecasting} proposed also smoothing past prices. They define:
\begin{equation}
l(t, k) 
\,=\, \frac{m_+(t, k) \;-\; m_-(t, k)}{m_-(t, k)} 
\quad\text{where}
\end{equation}
\begin{equation}
m_+\bigl(t, k\bigr) \;=\; \frac{1}{k+1}\sum_{i=0}^k p\bigl(t + i\bigr)\quad\text{and}
\end{equation}
\begin{equation}
m_-\bigl(t, k\bigr) \;=\; \frac{1}{k+1}\sum_{i=0}^k p\bigl(t - i\bigr),
\end{equation}
noting that $i$ runs from $0$ to $k$, so there are $(k+1)$ terms in the sum. A key drawback is that the window length $k$ coincides with the prediction horizon $h$. This can bias the labels: for instance, a horizon of $h=2$ may not provide enough smoothing, whereas a large horizon might over-smooth price moves.

To overcome this, we propose a more general labeling strategy that dissociates $k$ from $h$. Specifically, we define:
\begin{equation}
w_+\bigl(t, h, k\bigr) = \frac{1}{k+1}\,\sum_{i=0}^k p\bigl(t + h - i\bigr)
\end{equation}
\begin{equation}
w_-\bigl(t, h, k\bigr) = \frac{1}{k+1}\,\sum_{i=0}^k p\bigl(t - i\bigr).
\end{equation}
The percentage change is then
\begin{equation}
l\bigl(t, h, k\bigr) 
\,=\, \frac{\,w_+\bigl(t, h, k\bigr) \;-\; w_-\bigl(t, h, k\bigr)\,}
             {\,w_-\bigl(t, h, k\bigr)\!}.
\end{equation}
We classify a trend as \emph{upward} if $l(t,h,k) > \theta$, 
\emph{downward} if $l(t,h,k) < -\theta$, 
and \emph{stable} if $-\theta \,\leq\, l(t,h,k) \,\leq\, \theta$.
The threshold $\theta$ is often chosen to balance the three classes rather than to reflect trading costs. We argue, however, that relating $\theta$ to transaction costs can better align trend predictions with profitability. 
Thus, in Section~\ref{sec:theta}, we examine setting $\theta$ to the average spread (the difference between the best bid and ask prices) as a percentage of the mid-price\footnote{Expressing the spread as a percentage of the mid-price preserves consistency with $l(t,h,k)$, which is also a percentage.}, since the spread represents the main transaction cost.

Figure~\ref{fig:labeling} illustrates all three approaches. 
For a fair comparison with existing literature, we adopt the original labeling method in our FI-2010 experiments and use our new labeling strategy for Intel and Tesla data, where the more general approach better handles varying horizons.

\begin{figure*}[h!]
    \centering
    \includegraphics[width=\linewidth]{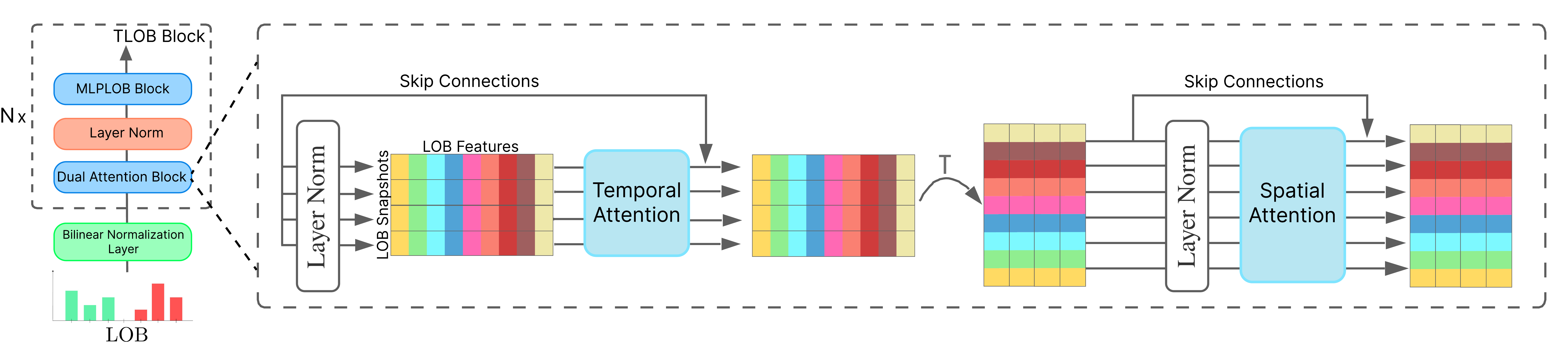}
    \caption{TLOB architecture overview. The model leverages Temporal Self-Attention and Feature Self-Attention within each TLOB block to capture time-wise and spatial relationships in Limit Order Book data. Each block is preceded by Bilinear Normalization to address non-stationarity, followed by an MLPLOB block.}
    \label{fig:architecture}
\end{figure*}
\section{Models}
\label{sec:models}
We propose two novel deep learning models for Price Trend Prediction (PTP) using Limit Order Book (LOB) data. The first, called \textbf{MLPLOB}, is a simple MLP-based model. The second, \textbf{TLOB}, leverages a dual-attention Transformer-based approach. Both models take as input a sequence of LOB time series consisting of the last $T$ LOB snapshots for 10 LOB levels.
\subsection{MLPLOB}
A key finding from the benchmark study by Prata et al.~\cite{prata2024lob} reveals that, despite the proliferation of specialized deep learning architectures for PTP, their performance often converges toward low values when tested on diverse and complex datasets. Inspired by the work of Tolstikhin et al.~\cite{tolstikhin2021mlp} and Zeng et al. ~\cite{zeng2023transformers}, who demonstrated that simple MLP-based models can perform as well as state-of-the-art (SoTA) methods in certain domains, we develop an MLP-based architecture for PTP with LOB data, called \emph{MLPLOB}.

\textbf{Architecture Overview.}
MLPLOB is composed of multiple blocks, each containing two types of MLP layers: (1.) \emph{Feature-Mixing MLPs}, which operate along the feature axis. (2.) \emph{Temporal-Mixing MLPs}, which operate along the time axis.
This design aims to capture both spatial and temporal relationships in LOB data--characteristics that Sirignano and Cont \cite{sirignano2019deep, sirignano2021universal} identified as fundamental to LOB dynamics and modeling.

Each MLP layer consists of two fully connected layers, mirroring the MLP component used in Transformer architectures~\cite{vaswani2017attention}. Initially, the input sequence is projected linearly into a tensor $\mathbf{X}\in \mathbb{R}^{T \times N}$, where $N$ is a chosen hyperparameter.

\textbf{Feature-Mixing MLPs.}
We apply a feature-mixing MLP row by row (\emph{i.e.}, for each time step $i$). Formally,
\begin{equation}
    \mathbf{U}_{i,*} = 
    \sigma\Bigl(
      \text{LayerNorm}\bigl(
          \sigma(\mathbf{X}_{i,*}\,\mathbf{W}_1)\,\mathbf{W}_2 
          \;+\; \mathbf{X}_{i,*}
      \bigr)
    \Bigr)
    \quad i=1,\dots,T,
\end{equation}
where $\sigma$ is the GeLU activation function~\cite{hendrycks2016gaussian}, and \(\text{LayerNorm}\) denotes layer normalization.

\textbf{Temporal-Mixing MLPs.}
Next, we transpose the resulting tensor $\mathbf{U}$ and apply a temporal-mixing MLP column by column (\emph{i.e.}, for each feature dimension $j$):
\begin{equation}
    \mathbf{Z}_{*,j} =
    \sigma\Bigl(
      \text{LayerNorm}\bigl(
          \sigma(\mathbf{U}_{*,j}\,\mathbf{W}_3)\,\mathbf{W}_4 
          \;+\; \mathbf{U}_{*,j}
      \bigr)
    \Bigr)
    \quad j=1,\dots,N.
\end{equation}

\textbf{Model Simplicity and Isotropic Design.}
The MLPLOB architecture relies only on matrix multiplications, reshaping operations, and scalar nonlinearities. It also adopts an \emph{isotropic design}, wherein each block (beyond the initial projection) has a constant dimensionality. This contrasts with the pyramidal layouts found in many CNNs (which reduce spatial resolution while increasing channel depth). Notably, isotropic designs are also common in Transformers and Recurrent Neural Networks (RNNs).

\textbf{Final Prediction.}
After several blocks of feature and temporal mixing, MLPLOB performs dimensionality reduction to collapse all features into a single vector, which then passes through several fully connected layers that gradually diminish the vector dimension and a final standard classification head. The network outputs the directional trend (up, down, or stable) for the final time step. Our primary objective in devising MLPLOB is to show that a carefully structured MLP-based model can match or exceed more complex architectures in the PTP task. The same method is also applied to TLOB. 
\subsection{TLOB}
The Transformer architecture~\cite{vaswani2017attention} has led to major breakthroughs in deep learning, notably in natural language processing~\cite{brown2020language, khan2022transformers} and time-series modeling~\cite{wen2022transformers}. A key advantage is the ability to capture long-range dependencies without suffering as much from vanishing gradients or forgetting, and performance typically scales favorably with increased data~\cite{kaplan2020scaling}. Because massive volumes of financial LOB data are available and long-range dependencies are central in predicting the price trend, Transformers are well-positioned for LOB modeling.

\textbf{Dual-Attention Blocks.}
We propose \emph{TLOB}, a Transformer-based architecture specifically designed for Limit Order Book data. Each TLOB block contains:
\begin{enumerate}
    \item \emph{Self-Attention over LOB Snapshots (Temporal Axis)}, computes attention values between different LOB snapshots, capturing time-wise dependencies among consecutive snapshots.
    \item \emph{Self-Attention over LOB Features (Spatial Axis)}, computes attention values between LOB features, capturing spatial relationships among different price-volume features.
    \item An \emph{MLPLOB block}, which replaces the usual Transformer feed-forward network to enhance the model’s capacity for combining spatial and temporal signals.
\end{enumerate}
The architecture is shown in Fig. \ref{fig:architecture}.

\textbf{Temporal vs. Feature Attention.}
While standard Transformers~\cite{vaswani2017attention} process tokens along a single dimension, LOB data naturally requires both temporal and spatial dependencies to be learned \cite{sirignano2019deep, sirignano2021universal}. For instance, time-step $t$ can reveal how deeper or shallower levels relate to one another, as well as how trends evolve over past snapshots. Hence, \emph{dual-attention} explicitly addresses these two axes of variation. To investigate the importance of each type of attention layers we performed an ablation study (Section \ref{sec:ablation}).

\textbf{Bilinear Normalization Layer.}
To address non-stationarity and magnitude disparity (prices and sizes) in financial time series, we employ a Bilinear Normalization layer~\cite{tran2021data} as the initial layer. Unlike conventional $z$-score normalization, which can fail under distribution shifts at inference time, bilinear normalization adapts to batch-specific statistics, maintaining robust performance even when market conditions change. The same layer is also used in MLPLOB.

\textbf{Positional Encoding.}
Because self-attention is permutation-invariant, we incorporate sinusoidal positional embeddings~\cite{vaswani2017attention} to preserve the chronological structure within each LOB window. This embedding ensures that TLOB respects the temporal ordering of snapshots, which is crucial for modeling price evolution.

By blending two distinct self-attention operations (temporal first, then spatial) with an MLPLOB feed-forward component, TLOB is designed to capture the complex market microstructure present in LOB data. Its Transformer foundation enables effective scaling for large datasets, while the dual-attention mechanism better handles the fine-grained feature interactions and sequence dependencies characteristic of financial time series.

\section{Experiments}
We conduct a comprehensive evaluation of MLPLOB and TLOB model training and testing on both the Benchmark FI-2010, the TSLA-INTC, and the BTC datasets. TLOB and MLPLOB surpass SoTA performances on every dataset and every horizon. TLOB performs the best on larger horizons, while MLPLOB performs the best on the shorter ones.
Our experiments extend beyond merely demonstrating the state-of-the-art performance of TLOB, aiming to address several critical research questions: (1) Are stock prices harder to forecast than in the past? (2) What if we choose $\theta$ equal to the average spread? (3) Are temporal and spatial attention necessary?
Through these investigations, we seek not only to validate our models' predictive capabilities but also to contribute to the broader understanding of deep learning applications in financial forecasting. 
\begin{table*}[h!]
\centering
\captionsetup{width=0.8\textwidth}
\caption{Intel and Tesla main characteristics for January 2015. Average liquidity is computed as the average quantity available in the first 10 LOB levels.} 
\label{tab:stocks} 
\begin{tabular*}{0.8\textwidth}{@{\extracolsep{\fill}} l c c c c @{}}
\toprule
\textbf{Stock} & \multicolumn{1}{c}{\textbf{Daily Return (\%) }} & \multicolumn{1}{c}{\textbf{Daily Volume}} & \textbf{Avg. Spread} & \textbf{Avg. Liquidity} \\
\midrule
TSLA & $-0.42 \pm 2.84$ & $23,927,602 \pm 4,554,884$ & $0.16$ & $3,320$ \\
INTC & $-0.44 \pm 1.66$ & $304,325,400 \pm 69,340,430$ & $0.01$ & $124,960$ \\
\bottomrule
\end{tabular*}
\end{table*}

\subsection{Benchmark dataset FI-2010}
\label{sec:fi-2010}
The FI-2010 dataset \cite{urn:nbn:fi:csc-kata20170601153214969115} is the most widely adopted LOB dataset within the field of deep learning applications to limit order books \cite{zhang2019deeplob, zhang2020deep, tsantekidis2017forecasting, tsantekidis2017using}, particularly for forecasting endeavors. It comprises LOB data from five Finnish companies listed on the NASDAQ Nordic stock exchange: Kesko Oyj, Outokumpu Oyj, Sampo, Rautaruukki, and Wärtsilä Oyj. The data span ten trading days, from June 1st to June 14th, 2010, encompassing approximately 4 million limit order snapshots across ten levels of the LOB. The authors sampled LOB observations at intervals of ten events, resulting in a total of 394,337 samples.
The dataset is pre-processed, with the labels already computed. 
The label associated with each data point, indicative of mid-price movement, is determined by the percentage change between the current mid-price and the average of the subsequent $h$ (chosen horizon) mid-prices. The percentage change is thus defined as:
\begin{equation}
l(t) = \frac{m_+(t, k) - p(t)}{p(t)} 
\end{equation}
where $p(t)$ is the mid-price and $k$ represents the window length, which in this instance also corresponds to the prediction horizon $h$. Labels are assigned as explained in \ref{sec:task}. The dataset furnishes time series and corresponding class labels for five distinct horizons: $h \in H = \{10, 20, 30, 50, 100\}$. The dataset's authors employed a uniform threshold $\theta = 2\times 10^{-3}$ across all horizons. The value is chosen to balance the classes for $h = 50$. 

\subsection{BTC Dataset}
The BTC dataset is extracted from Kaggle and encompasses 12 consecutive days, starting from January 9th, 2023, until January 20th, 2023. It comprises data from Binance Bitcoin perpetual data (BTCUSDT.P). The data points are already sampled at a frequency of 250 milliseconds, consequently, the four selected horizons (10, 20, 50, 100) correspond to a time horizon of 1s, 2.5s, 5s, and 10s, respectively. The dataset contains a total of 3.730.870 rows and it is partitioned such that the initial 9 days are allocated for training, the 10th day for validation, and the final day for testing. 

\subsection{TSLA-INTC Dataset}
In the majority of state-of-the-art (SoTA) research within the domain of Deep Learning applied to LOB data, researchers typically employ one, two, or three stocks \cite{coletta2021towards, nagy2023generative, li2020generating, shi2022state, hultin2023generative}, predominantly from the technology sector. Adhering to this established practice, we construct a LOB dataset comprising two NASDAQ-listed stocks, namely, Tesla and Intel -- spanning the period from January 2nd to January 30th, 2015. We posit that stylized facts and market microstructure characteristics exhibit independence from individual stock behaviors (as demonstrated in \cite{bouchaud2009markets, bouchaud2002statistical, cont2014price, gould2013limit}\footnote{These seminal works in finance elucidate the universal statistical properties of LOBs, transcending specific stocks and markets.}). The dataset encompasses 20 order book files for each stock, corresponding to each trading day, resulting in a total of approximately 24 million samples. 
The dataset is partitioned such that the initial 17 days are allocated for training, the 18th day for validation, and the final two days for testing. As shown in Table \ref{tab:stocks}, the main characteristics of Tesla and Intel for January 2015 are very different, offering different market situations. Differently from TSLA, INTC is a small tick stock. Unfortunately, we cannot make the dataset public for copyright reasons.

\textbf{Sampling}. Limit Order Book data, especially for liquid stocks, is massive, every day, hundreds of thousands of orders are placed for each stock. Furthermore, financial data are known to have a low signal-to-noise ratio \cite{nagel2021machine}. Accordingly, it is unnecessary to consider every LOB update, so defining a valid sampling technique is essential. While time-based\footnote{BTC dataset is sampled every 100ms.} and event-based sampling methods\footnote{FI-2010 is sampled every 10 events.} are used, they fail to capture the varying impact of transactions. In fact, single transactions can have very different impacts on the market. Volume-based sampling offers a solution by sampling the LOB after a predetermined volume of shares has been traded, thus reflecting the magnitude of market activity. Therefore, we adopted a sampling strategy based on trading volume, where snapshots of the Limit Order Book (LOB) are taken every 500 stocks traded. This method achieves a compromise between maintaining adequate temporal consistency within windows and ensuring significant variation between samples.

\subsection{Experimental settings}
For each dataset, we trained and tested the performance of each model on different horizons, namely $10$, $20$, $50$, and $100$\footnote{depending on the sampling method, the horizon is in a different unit of measure}. 
All the experiments were carried out using an RTX 3090.
Since the FI-2010 dataset also contains 104 handcrafted features derived from the LOB, we used them in both our models. This choice improved the performance of the F1-Score by approximately $1$.
For Tesla and Intel, given the availability of message files containing the order information, we augmented the LOB snapshots by concatenating them with the corresponding orders. This integration was undertaken to incorporate additional information not present in the LOB. Consequently, this approach resulted in an approximate improvement of $1.5$ in the F1-score. We report the details on the hyperparameter search in the Supplementary Material.

\textbf{Baselines} As comparative baselines, we employed 3 machine learning models: Support Vector Machine (SVM), Random Forest and XGBoost, and 10 deep learning SoTA LOB-based models: MLP, LSTM \cite{tsantekidis2017forecasting}, CNN \cite{tsantekidis2017using}, CTABL \cite{tran2018temporal}, DAIN \cite{passalis2019deep}, CNNLSTM \cite{tsantekidis2020using}, DeepLOB \cite{zhang2019deeplob}, BiN-CTABL \cite{tran2021data}, AXIALLOB \cite{10022284} and DLA \cite{guo2023forecasting}. Due to computational constraints, we selected the top two performing models from \cite{prata2024lob}, specifically DeepLOB and BiNCTABL, and exclusively trained and tested these models with the TSLA-INTC and BTC datasets. In Table \ref{tab:inference}, we report the inference time (ms) and the number of parameters for each SoTA model. Although TLOB and MLPLOB have a higher number of parameters compared to SoTA LOB-based models, they still have significantly fewer parameters than state-of-the-art 
\begin{table}[h!]
\centering
\begin{minipage}{0.48\textwidth}
\caption{Number of parameters and inference time for each model used in the experiments.} 
\label{tab:inference}
\centering
        \begin{tabular}{l|cc}
            \toprule
            \textbf{Model} & \textbf{Nr. parameters} & \textbf{Inference Time (ms)} \\
            \midrule
            MLP & $10^6$ & 0.08 \\
            LSTM \cite{tsantekidis2017using} & $1.6\cdot10^4$ & 0.21 \\
            CNN \cite{tsantekidis2017forecasting} & $3.5\cdot10^4$ & 0.36  \\
            CTABL \cite{tran2018temporal} & $1.1 \cdot 10^4$ & 0.48 \\
            DAIN-MLP \cite{passalis2019deep} & $5.3\cdot10^4$ & 0.50  \\
            CNNLSTM \cite{tsantekidis2020using} & $2.8 \cdot 10^5$ & 0.49  \\
            AXIALLOB \cite{10022284} & $2\cdot10^4$ & 1.91 \\
            DLA \cite{guo2023forecasting} & $1.2\cdot10^5$ & 0.23  \\
            DeepLOB \cite{zhang2019deeplob} & $1.4\cdot10^5 $& 1.31 \\
            BiNCTABL \cite{tran2021data} & $1.1 \cdot 10^4$ & 0.71  \\
            \midrule
            MLPLOB & $6.3 \cdot 10^7$ & 4.79 \\
            TLOB & $1 \cdot 10^7$ & 2.24 \\
            \bottomrule
        \end{tabular}
        \end{minipage}
\end{table}
deep learning models commonly used in standard machine learning tasks. Furthermore, they do result in slightly higher inference times, but their speed remains adequate for application in high-frequency trading scenarios.

\textbf{Trend Classification Threshold} We remark that $\theta$ is the parameter that determines if a percentage change $l_t$ is classified as an up, stable, or downtrend. For the TSLA-INTC and the BTC datasets, to ensure balanced class distribution, we set $\theta$ equal to the mean percentage change. In Sec. \ref{sec:theta} we explore an alternative approach to defining $\theta$ based on financial parameters rather than class balance optimization. For the FI-2010 dataset, we retained the original labels to maintain consistency with existing benchmark studies and previous works.

\textbf{Metric} We selected the F1-score as our primary performance metric because it captures both precision and recall in a single value. Accuracy is not a valid metric for our experiments because the classes are not balanced for each horizon. The F1-score is robust to the class imbalance problem, which detrimentally affects the accuracy. Finally, the F1-score is the most used metric in the SoTA papers tackling the PTP task. For a comprehensive evaluation, we provide precision and recall curves in the Supplementary Material.
\section{Results}

\begin{table}[h!]
\centering
\begin{minipage}{0.48\textwidth}
\caption{F1-score on the FI-2010 dataset on four horizons. 
Bold values show the best scores.} 
\label{tab:f1-2010}
\centering
        \begin{tabular}{l|cccc}
            \toprule
            & \multicolumn{4}{c}{\textbf{FI-2010 F1-Score (\%) }$\uparrow$} \\
            \cmidrule(lr){2-5} 
            \textbf{Model} & \textbf{h = 10} & \textbf{h = 20} & \textbf{h = 50} & \textbf{h = 100} \\
            \midrule
            SVM & 35.9 & 43.2 & 49.4 & 51.2 \\
            Random Forest & 48.7 & 46.3 & 51.2 & 53.9 \\
            XGBoost & 62.4 & 59.6 & 65.3 & 67.6 \\
            MLP & 48.2 & 44.0 & 49.0 & 51.6 \\
            LSTM \cite{tsantekidis2017using} & 66.5 & 58.8 & 66.9 & 59.4  \\
            CNN \cite{tsantekidis2017forecasting} & 49.3 & 46.1 & 65.8 & 67.2 \\
            CTABL \cite{tran2018temporal} & 69.5 & 62.4 &  71.6 & 73.9 \\
            DAIN-MLP \cite{passalis2019deep} & 53.9 & 46.7 & 61.2 & 62.8 \\
            CNNLSTM \cite{tsantekidis2020using} & 63.5 & 49.1 & 69.2 & 71.0 \\
            AXIALLOB \cite{10022284} & 73.2 & 63.4 & 78.3 & 79.2 \\
            DLA \cite{guo2023forecasting} & 79.4 & 69.3 & 87.1 & 52.2 \\
            DeepLOB \cite{zhang2019deeplob} & 71.1 & 62.4 & 75.4 & 77.6 \\
            BiNCTABL \cite{tran2021data} & 81.1 & 71.5 & 87.7 & 92.1 \\
            \midrule
            MLPLOB & \textbf{81.64} & \textbf{84.88} & \textbf{91.39} & 92.62 \\
            TLOB & 81.55 & 82.68 & 90.03 & \textbf{92.81} \\
            \bottomrule
        \end{tabular}
        \end{minipage}
\end{table}

\subsection{FI-2010 results}
Table \ref{tab:f1-2010} presents the performance comparison across four prediction horizons\footnote{Note that the horizon values represent the number of events before the sampling process of the dataset, while in the benchmarks \cite{prata2024lob, urn:nbn:fi:csc-kata20170601153214969115} the values represent the horizons after the sampling process. In other words, the horizons considered are the same and are the ones defined originally in FI-2010.} for the FI-2010 benchmark dataset. In the Supplementary Material, we also report the precision and recall curves for horizon 100. MLPLOB and TLOB exhibit very high precision, also at high recall values, and consistently achieve higher precision at all recall levels compared to the other models.
The results for the baselines are extracted from the benchmark of Prata et al. \cite{prata2024lob}\footnote{if we had taken the results reported in the individual papers, MLPLOB and TLOB would have still outperformed all the other models.} since the settings are equal for the FI-2010 dataset. 
MLPLOB and TLOB outperform all the SoTA LOB-based models, surpassing state-of-the-art performance. Notably, the performance differential between MLPLOB and TLOB is minimal, which, as we will demonstrate in Section \ref{sec:lobster}, can be attributed to the lower complexity of the FI-2010 dataset, which explains the uselessness of a more complex architecture such as TLOB for this particular dataset.

\subsection{Tesla and Intel results}
\label{sec:lobster}

\begin{table}[h!]
\centering
\begin{minipage}{0.48\textwidth}
\centering
\caption{F1-score for Tesla on four horizons. Bold values show the best scores.} 
\label{tab:tesla}
\centering
        \begin{tabular}{c|cccc}
            \toprule
            & \multicolumn{4}{c}{\textbf{TSLA F1-Score (\%) }$\uparrow$} \\
            \cmidrule(lr){2-5} 
            \textbf{Model} & \textbf{h = 10} & \textbf{h = 20} & \textbf{h = 50} & \textbf{h = 100}  \\
            \midrule
            DeepLOB & 36.25 & 36.58 & 35.29 & 34.43 \\
            BiNCTABL & 58.69 & 48.83 & 42.23 & 38.77 \\
            \midrule
            MLPLOB & \textbf{60.72} & \textbf{50.25} & 38.97 & 32.95 \\
            TLOB & 60.50 & 49.74 & \textbf{43.48} & \textbf{39.84}   \\            \bottomrule
        \end{tabular}
        \end{minipage}
\end{table}

\begin{table}[h!]
\centering
\begin{minipage}{0.48\textwidth}
\centering
\caption{F1-score for Intel on four horizons. Bold values show the best scores.} 
\label{tab:intel}
\centering
        \begin{tabular}{c|cccc}
            \toprule
            & \multicolumn{4}{c}{\textbf{INTC F1-Score (\%) }$\uparrow$} \\
            \cmidrule(lr){2-5} 
            \textbf{Model} & \textbf{h = 10} & \textbf{h = 20} & \textbf{h = 50} & \textbf{h = 100}  \\
            \midrule
            DeepLOB & 68.13 & 63.70 & 40.3 & 30.1 \\
            BiNCTABL & 72.65 & 66.57 & 53.99 & 41.08 \\
            \midrule
            MLPLOB & \textbf{81.15} & \textbf{73.25} & 55.74 & 43.18 \\
            TLOB & 80.15 & 72.75 & \textbf{62.07} & \textbf{50.14} \\
            \bottomrule
        \end{tabular}
        \end{minipage}
\end{table}

In Table \ref{tab:tesla} we show the results for Tesla and in Table \ref{tab:intel} for Intel.
For each stock, we trained a different model. 
In the Supplementary Material, we also report the precision and recall curves for a horizon equal to 100. For INTC, they exhibit excellent precision at low recall values, indicating their ability to accurately identify the most confident positive instances.
MLPLOB outperforms every model on the first two horizons (10, 20), while on the longer horizons (50, 100), TLOB outperforms every model. This is expected since Transformers excels at long-range dependencies. 
Notably, the difference in performance between MLPLOB and TLOB for the shorter horizons is minimal ($\approx 0.5$), while on the longer horizons it is significant ($\approx 7$). As expected, the longer the horizon, the more difficult to forecast.
In general, the performances are much lower with respect to FI-2010. We conjecture that this is due to the fact that FI-2010 is characterized by a lower level of complexity with respect to NASDAQ stocks. This derives from the fact that it is composed of Finnish stocks, which are less liquid and efficient than NASDAQ stocks such as Intel and Tesla. Additionally, the data of FI-2010 dates back to 2010. Indeed, as will be demonstrated in the subsequent experiment, the prediction difficulty augments as time goes by. 
All the models are trained until convergence. Notably, both TLOB and MLPLOB achieve convergence in less than half the epochs required by BiNCTABL and DeepLOB.

\subsection{BTC Results}
In Table \ref{tab:btc} we show the results for the BTC dataset. We remark that this dataset is from 2023, so it is the most recent one \footnote{FI-2010 is from 2010, while TSLA-INTC is from 2015}.
Unlike the other datasets, TLOB outperforms every model on every horizon. This consistent dominance on the most recent (2023) dataset, particularly with a volatile asset like Bitcoin, suggests TLOB's architecture is highly effective at capturing contemporary market dynamics that may elude other models. Moreover, the widening performance gap at longer horizons (h=50, 100) with respect to MLPLOB hints at TLOB's enhanced capacity to model the complex, longer-term temporal dependencies.
Similarly to the other two datasets, the difference in performance between MLPLOB and TLOB for the shorter horizons (10, 20) is minimal ($\approx 0.5$), while on the longer horizons it is significant ($\approx 5$). 
\begin{table}[h!]
\centering
\begin{minipage}{0.48\textwidth}
\centering
\caption{F1-score for BTC on four horizons. Bold values show the best scores.} 
\label{tab:btc}
\centering
        \begin{tabular}{c|cccc}
            \toprule
            & \multicolumn{4}{c}{\textbf{BTC F1-Score (\%) }$\uparrow$} \\
            \cmidrule(lr){2-5} 
            \textbf{Model} & \textbf{h = 10} & \textbf{h = 20} & \textbf{h = 50} & \textbf{h = 100}  \\
            \midrule
            DeepLOB & 68.07 & 57.87 & 45.13 & 37.43 \\
            BiNCTABL & 73.4 & 61.34 & 47.05 & 40.59 \\
            \midrule
            MLPLOB & 74.6 & 61.02 & 42.74 & 36.97 \\
            TLOB & \textbf{74.7} & 61.74 & \textbf{48.54} & \textbf{41.49}   \\            \bottomrule
        \end{tabular}
        \end{minipage}
\end{table}

\subsection{Are stocks harder to forecast than in the past?}
\begin{table}[h!]
\centering
\begin{minipage}{0.48\textwidth}
\centering
\caption{F1-score for Intel on two different periods, from 2012 and 2015. The horizon is set to 50.} 
\label{tab:intel2}
\centering
        \begin{tabular}{c|cc}
            \toprule
            & \multicolumn{2}{c}{\textbf{F1-Score (\%) }$\uparrow$} \\
            \cmidrule(lr){2-3} 
            \textbf{Model} & \textbf{INTC 2015} & \textbf{INTC 2012} \\
            \midrule
            TLOB & 60.19 & 66.87 \\
            \bottomrule
        \end{tabular}
        \end{minipage}
\end{table}
\label{sec:past}
This experiment examines the challenges associated with market prediction over time and the self-destruction of predictable patterns in financial markets. Empirical evidence consistently demonstrates that forecasting models effective in certain periods become obsolete over time. Several studies indicate that previously observed predictability patterns disappeared after becoming widely known. Dimson and Marsh \cite{dimson1999murphy} found this for the UK small-cap premium, while Bossaert and Hillion \cite{bossaerts1999implementing} noted a decline in international stock return predictability around 1990. Aiolfi and Favero \cite{aiolfi2005model} reported similar findings for US stocks in the 1990s. The market is increasingly efficient and difficult to predict as time goes by. We extend this investigation to our best-performing model, TLOB. Specifically, we tested on a day of Intel from 2012/06/21\footnote{we remark that in a single day of Intel, there are hundreds of thousands of order,s making the experiment statistically significant. Furthermore, the trading day was extracted from the LOBSTER public sample files available at \url{https://lobsterdata.com/info/DataSamples.php}, and it was the only day available, eliminating the possibility of cherry picking.} and confronted the difference in performance with  2015/01/30. We report the performance in Table \ref{tab:intel2}. As expected, the performance from 2012 is better than that from 2015. We confirm the hypothesis and the empirical evidence from other works.

\subsection{Alternative Threshold Definition Using Average Spread}
\begin{table}[h!]
\centering
\begin{minipage}{0.48\textwidth}
\centering
        \caption{F1-score on Tesla with $\theta$ set to the average spread.} 
        \label{tab:theta}
        \begin{tabular}{c|ccc}
            \toprule
            & \multicolumn{3}{c}{\textbf{F1-Score (\%) }$\uparrow$} \\
            \cmidrule(lr){2-4} 
            \textbf{Model} & \textbf{h = 50} & \textbf{h = 100} & \textbf{h = 200}  \\
            \midrule
            TLOB & 41.39 & 36.48 & 30.82 \\
            \bottomrule
        \end{tabular}
        \end{minipage}
\end{table}
\label{sec:theta}

Based on the fact that predictability has to be considered in relation to the transaction costs, we explore an alternative approach to define the trend classification parameter $\theta$, setting it equal to the average spread as a percentage of the mid-price, reflecting the primary transaction cost. 
This methodology could only be applied to Tesla data, as Intel's higher trading volume (approximately 10 times greater in January 2015) and lower volatility relative to traded shares would result in 99.99\% of trends classified as stationary.
We set the horizons to 50, 100, and 200 because with shorter horizons, 99\% of the mid-price movements would be classified as stationary. 
In Table \ref{tab:theta} we report the results. 
In general, performances show a deterioration, which is probably caused by the classes' unbalance.
This experiment highlights the necessity for further refinements in trend definition and method complexity when targeting profitability in practical applications.

\subsection{Ablation Study}
\label{sec:ablation}
To evaluate the contribution of each attention mechanism within the TLOB architecture, we performed an ablation study on the FI-2010 dataset. Specifically, we compared the performance of the complete TLOB model against two ablated versions: one without spatial attention (TLOB w/o SA) and another without temporal attention (TLOB w/o TA). To avoid inconsistency, we maintain the total number of layers fixed\footnote{TLOB has 4 temporal attention layers and 4 spatial attention layers, TLOB w/o SA has 8 temporal attention layers, and TLOB w/o TA has 8 spatial attention layers}. The F1-scores for each model across four prediction horizons (h = 10, 20, 50, and 100) are presented in Table \ref{tab:ablation}. The results demonstrate that the full TLOB model, incorporating both spatial and temporal attention mechanisms, consistently outperforms both ablated versions across all prediction horizons. The performance gain of the full TLOB model highlights the importance of capturing both spatial relationships between LOB features and temporal dependencies across LOB snapshots. This suggests that the dual-attention mechanism effectively learns complementary information, leading to improved predictive accuracy compared to models relying on only one type of attention.

\vspace{0.3cm}
\begin{table}
\centering
\begin{minipage}{0.48\textwidth}
\caption{Ablation study results. F1-score on the FI-2010 dataset on four horizons. 
Bold values show the best scores.} 
\label{tab:ablation}
\centering
        \begin{tabular}{l|cccc}
            \toprule
            & \multicolumn{4}{c}{\textbf{FI-2010 F1-Score (\%) } $\uparrow$} \\
            \cmidrule(lr){2-5} 
            \textbf{Model} & \textbf{h = 10} & \textbf{h = 20} & \textbf{h = 50} & \textbf{h = 100} \\
            \midrule
            TLOB w/o SA & 79.59 & 78.96 &  87.51 & 91.40 \\
            TLOB w/o TA & 80.27 & 79.20  & 87.72 & 91.42 \\
            \midrule
            TLOB & \textbf{81.55} & \textbf{82.68} & \textbf{90.03} & \textbf{92.81} \\
            \bottomrule
        \end{tabular}
        \end{minipage}
\end{table}
\vspace{-0.3cm}
\section{Conclusion}
We proposed MLPLOB and TLOB, two deep‑learning architectures for LOB-based price trend prediction. Both outperformed SoTA methods on FI‑2010 (Finnish stocks), NASDAQ stocks (Tesla, Intel), and Bitcoin, especially for longer horizons.
When considering practical implementation, we found that defining trend thresholds based on average spread (transaction costs) significantly impacts model performance and profitability. This finding underscores the critical gap between machine learning metrics and practical trading applicability.

\textbf{Future works}: Looking ahead, several avenues for future research emerge. The investigation of scaling laws for financial deep learning models remains an open question, as does the development of more robust approaches to handling increased market efficiency and complexity. Additionally, the exploration of alternative trend definition methodologies that better align with practical trading constraints could prove fruitful. Finally an extensive profitability analysis, based on backtesting or other more advanced market simulations methods \cite{coletta2021towards, coletta2022learning, berti2025trades}, would be very interesting.

\textbf{Limitations}: Firstly, it is important to acknowledge that the proposed methodologies are not sufficiently mature for practical deployment in live trading environments. When considering practical implementation, we found that defining trend thresholds based on average spread (transaction costs) significantly impacts model evaluation and potential profitability. This finding underscores the critical gap between academic performance metrics and practical trading applicability.

\textbf{Risks}: Automated ML models, increasingly integrated into financial markets, present significant risks to financial stability due to their potential to amplify systemic vulnerabilities. These models can trigger rapid and widespread market reactions, exacerbating market volatility and potentially leading to cascading failures across the financial system.

\newpage
\bibliography{bibliography}

\begin{thebibliography}{46}
\providecommand{\natexlab}[1]{#1}
\providecommand{\url}[1]{\texttt{#1}}
\expandafter\ifx\csname urlstyle\endcsname\relax
  \providecommand{\doi}[1]{doi: #1}\else
  \providecommand{\doi}{doi: \begingroup \urlstyle{rm}\Url}\fi

\bibitem[Aiolfi and Favero(2005)]{aiolfi2005model}
M.~Aiolfi and C.~A. Favero.
\newblock Model uncertainty, thick modelling and the predictability of stock returns.
\newblock \emph{Journal of Forecasting}, 24\penalty0 (4):\penalty0 233--254, 2005.

\bibitem[Berti et~al.(2025)Berti, Prenkaj, and Velardi]{berti2025trades}
L.~Berti, B.~Prenkaj, and P.~Velardi.
\newblock Trades: Generating realistic market simulations with diffusion models.
\newblock \emph{arXiv preprint arXiv:2502.07071}, 2025.

\bibitem[Bossaerts and Hillion(1999)]{bossaerts1999implementing}
P.~Bossaerts and P.~Hillion.
\newblock Implementing statistical criteria to select return forecasting models: what do we learn?
\newblock \emph{The Review of Financial Studies}, 12\penalty0 (2):\penalty0 405--428, 1999.

\bibitem[Bouchaud et~al.(2018)Bouchaud, Bonart, Donier, and Gould]{bouchaud2018trades}
J.~Bouchaud, J.~Bonart, J.~Donier, and M.~Gould.
\newblock \emph{Trades, Quotes and Prices: Financial Markets Under the Microscope}.
\newblock Cambridge University Press, 2018.
\newblock ISBN 9781107156050.
\newblock URL \url{https://books.google.it/books?id=u45LDwAAQBAJ}.

\bibitem[Bouchaud et~al.(2002)Bouchaud, M{\'e}zard, and Potters]{bouchaud2002statistical}
J.-P. Bouchaud, M.~M{\'e}zard, and M.~Potters.
\newblock Statistical properties of stock order books: empirical results and models.
\newblock \emph{Quantitative finance}, 2\penalty0 (4):\penalty0 251, 2002.

\bibitem[Bouchaud et~al.(2009)Bouchaud, Farmer, and Lillo]{bouchaud2009markets}
J.-P. Bouchaud, J.~D. Farmer, and F.~Lillo.
\newblock How markets slowly digest changes in supply and demand.
\newblock In \emph{Handbook of financial markets: dynamics and evolution}, pages 57--160. Elsevier, 2009.

\bibitem[Brown(2020)]{brown2020language}
T.~B. Brown.
\newblock Language models are few-shot learners.
\newblock \emph{arXiv preprint arXiv:2005.14165}, 2020.

\bibitem[Byrd et~al.(2020)Byrd, Hybinette, and Balch]{byrd2020abides}
D.~Byrd, M.~Hybinette, and T.~H. Balch.
\newblock Abides: Towards high-fidelity multi-agent market simulation.
\newblock In \emph{Proceedings of the 2020 ACM SIGSIM Conference on Principles of Advanced Discrete Simulation}, pages 11--22, 2020.

\bibitem[Chen et~al.(2024)Chen, Liang, Huang, Real, Wang, Pham, Dong, Luong, Hsieh, Lu, et~al.]{chen2024symbolic}
X.~Chen, C.~Liang, D.~Huang, E.~Real, K.~Wang, H.~Pham, X.~Dong, T.~Luong, C.-J. Hsieh, Y.~Lu, et~al.
\newblock Symbolic discovery of optimization algorithms.
\newblock \emph{Advances in neural information processing systems}, 36, 2024.

\bibitem[Coletta et~al.(2021)Coletta, Prata, Conti, Mercanti, Bartolini, Moulin, Vyetrenko, and Balch]{coletta2021towards}
A.~Coletta, M.~Prata, M.~Conti, E.~Mercanti, N.~Bartolini, A.~Moulin, S.~Vyetrenko, and T.~Balch.
\newblock Towards realistic market simulations: a generative adversarial networks approach.
\newblock In \emph{Proceedings of the Second ACM International Conference on AI in Finance}, pages 1--9, 2021.

\bibitem[Coletta et~al.(2022)Coletta, Moulin, Vyetrenko, and Balch]{coletta2022learning}
A.~Coletta, A.~Moulin, S.~Vyetrenko, and T.~Balch.
\newblock Learning to simulate realistic limit order book markets from data as a world agent.
\newblock In \emph{Proceedings of the Third ACM International Conference on AI in Finance}, pages 428--436, 2022.

\bibitem[Cont(2001)]{cont2001empirical}
R.~Cont.
\newblock Empirical properties of asset returns: stylized facts and statistical issues.
\newblock \emph{Quantitative finance}, 1\penalty0 (2):\penalty0 223, 2001.

\bibitem[Cont(2011)]{cont2011statistical}
R.~Cont.
\newblock Statistical modeling of high-frequency financial data.
\newblock \emph{IEEE Signal Processing Magazine}, 28\penalty0 (5):\penalty0 16--25, 2011.

\bibitem[Cont et~al.(2014)Cont, Kukanov, and Stoikov]{cont2014price}
R.~Cont, A.~Kukanov, and S.~Stoikov.
\newblock The price impact of order book events.
\newblock \emph{Journal of financial econometrics}, 12\penalty0 (1):\penalty0 47--88, 2014.

\bibitem[Diederik(2014)]{diederik2014adam}
P.~K. Diederik.
\newblock Adam: A method for stochastic optimization.
\newblock \emph{(No Title)}, 2014.

\bibitem[Dimson and Marsh(1999)]{dimson1999murphy}
E.~Dimson and P.~Marsh.
\newblock Murphy's law and market anomalies.
\newblock \emph{Journal of Portfolio Management}, 25\penalty0 (2):\penalty0 53--69, 1999.

\bibitem[Gould et~al.(2013)Gould, Porter, Williams, McDonald, Fenn, and Howison]{gould2013limit}
M.~D. Gould, M.~A. Porter, S.~Williams, M.~McDonald, D.~J. Fenn, and S.~D. Howison.
\newblock Limit order books.
\newblock \emph{Quantitative Finance}, 13\penalty0 (11):\penalty0 1709--1742, 2013.

\bibitem[Guo and Chen(2023)]{guo2023forecasting}
Y.~Guo and X.~Chen.
\newblock Forecasting the mid-price movements with high-frequency lob: a dual-stage temporal attention-based deep learning architecture.
\newblock \emph{Arabian Journal for Science and Engineering}, 48\penalty0 (8):\penalty0 9597--9618, 2023.

\bibitem[Hendrycks and Gimpel(2016)]{hendrycks2016gaussian}
D.~Hendrycks and K.~Gimpel.
\newblock Gaussian error linear units (gelus).
\newblock \emph{arXiv preprint arXiv:1606.08415}, 2016.

\bibitem[Hultin et~al.(2023)Hultin, Hult, Proutiere, Samama, and Tarighati]{hultin2023generative}
H.~Hultin, H.~Hult, A.~Proutiere, S.~Samama, and A.~Tarighati.
\newblock A generative model of a limit order book using recurrent neural networks.
\newblock \emph{Quantitative Finance}, pages 1--28, 2023.

\bibitem[Kaplan et~al.(2020)Kaplan, McCandlish, Henighan, Brown, Chess, Child, Gray, Radford, Wu, and Amodei]{kaplan2020scaling}
J.~Kaplan, S.~McCandlish, T.~Henighan, T.~B. Brown, B.~Chess, R.~Child, S.~Gray, A.~Radford, J.~Wu, and D.~Amodei.
\newblock Scaling laws for neural language models.
\newblock \emph{arXiv preprint arXiv:2001.08361}, 2020.

\bibitem[Khan et~al.(2022)Khan, Naseer, Hayat, Zamir, Khan, and Shah]{khan2022transformers}
S.~Khan, M.~Naseer, M.~Hayat, S.~W. Zamir, F.~S. Khan, and M.~Shah.
\newblock Transformers in vision: A survey.
\newblock \emph{ACM computing surveys (CSUR)}, 54\penalty0 (10s):\penalty0 1--41, 2022.

\bibitem[Kisiel and Gorse(2022{\natexlab{a}})]{10022284}
D.~Kisiel and D.~Gorse.
\newblock Axial-lob: High-frequency trading with axial attention.
\newblock In \emph{2022 IEEE Symposium Series on Computational Intelligence (SSCI)}, pages 1327--1333, 2022{\natexlab{a}}.
\newblock \doi{10.1109/SSCI51031.2022.10022284}.

\bibitem[Kisiel and Gorse(2022{\natexlab{b}})]{kisiel2022axial}
D.~Kisiel and D.~Gorse.
\newblock Axial-lob: High-frequency trading with axial attention.
\newblock In \emph{2022 IEEE Symposium Series on Computational Intelligence (SSCI)}, pages 1327--1333. IEEE, 2022{\natexlab{b}}.

\bibitem[Kissell(2020)]{kissell2020algorithmic}
R.~Kissell.
\newblock \emph{Algorithmic trading methods: Applications using advanced statistics, optimization, and machine learning techniques}.
\newblock Academic Press, 2020.

\bibitem[Li et~al.(2020)Li, Wang, Lin, Sinha, and Wellman]{li2020generating}
J.~Li, X.~Wang, Y.~Lin, A.~Sinha, and M.~Wellman.
\newblock Generating realistic stock market order streams.
\newblock In \emph{Proceedings of the AAAI Conference on Artificial Intelligence}, volume~34, pages 727--734, 2020.

\bibitem[Nagel(2021)]{nagel2021machine}
S.~Nagel.
\newblock \emph{Machine learning in asset pricing}, volume~1.
\newblock Princeton University Press, 2021.

\bibitem[Nagy et~al.(2023)Nagy, Frey, Sapora, Li, Calinescu, Zohren, and Foerster]{nagy2023generative}
P.~Nagy, S.~Frey, S.~Sapora, K.~Li, A.~Calinescu, S.~Zohren, and J.~Foerster.
\newblock Generative ai for end-to-end limit order book modelling: A token-level autoregressive generative model of message flow using a deep state space network.
\newblock \emph{arXiv preprint arXiv:2309.00638}, 2023.

\bibitem[Ntakaris et~al.()Ntakaris, Magris, Kanniainen, Gabbouj, and Iosifidis]{urn:nbn:fi:csc-kata20170601153214969115}
A.~Ntakaris, M.~Magris, J.~Kanniainen, M.~Gabbouj, and A.~Iosifidis.
\newblock Benchmark dataset for mid-price forecasting of limit order book data with machine learning methods.
\newblock \url{http://urn.fi/urn:nbn:fi:csc-kata20170601153214969115}.
\newblock N/A.

\bibitem[Passalis et~al.(2019)Passalis, Tefas, Kanniainen, Gabbouj, and Iosifidis]{passalis2019deep}
N.~Passalis, A.~Tefas, J.~Kanniainen, M.~Gabbouj, and A.~Iosifidis.
\newblock Deep adaptive input normalization for time series forecasting.
\newblock \emph{IEEE transactions on neural networks and learning systems}, 31\penalty0 (9):\penalty0 3760--3765, 2019.

\bibitem[Prata et~al.(2024)Prata, Masi, Berti, Arrigoni, Coletta, Cannistraci, Vyetrenko, Velardi, and Bartolini]{prata2024lob}
M.~Prata, G.~Masi, L.~Berti, V.~Arrigoni, A.~Coletta, I.~Cannistraci, S.~Vyetrenko, P.~Velardi, and N.~Bartolini.
\newblock Lob-based deep learning models for stock price trend prediction: a benchmark study.
\newblock \emph{Artificial Intelligence Review}, 57\penalty0 (5):\penalty0 1--45, 2024.

\bibitem[Shi and Cartlidge(2022)]{shi2022state}
Z.~Shi and J.~Cartlidge.
\newblock State dependent parallel neural hawkes process for limit order book event stream prediction and simulation.
\newblock In \emph{Proceedings of the 28th ACM SIGKDD Conference on Knowledge Discovery and Data Mining}, pages 1607--1615, 2022.

\bibitem[Sirignano and Cont(2021)]{sirignano2021universal}
J.~Sirignano and R.~Cont.
\newblock Universal features of price formation in financial markets: perspectives from deep learning.
\newblock In \emph{Machine learning and AI in finance}, pages 5--15. Routledge, 2021.

\bibitem[Sirignano(2019)]{sirignano2019deep}
J.~A. Sirignano.
\newblock Deep learning for limit order books.
\newblock \emph{Quantitative Finance}, 19\penalty0 (4):\penalty0 549--570, 2019.

\bibitem[Tolstikhin et~al.(2021)Tolstikhin, Houlsby, Kolesnikov, Beyer, Zhai, Unterthiner, Yung, Steiner, Keysers, Uszkoreit, et~al.]{tolstikhin2021mlp}
I.~O. Tolstikhin, N.~Houlsby, A.~Kolesnikov, L.~Beyer, X.~Zhai, T.~Unterthiner, J.~Yung, A.~Steiner, D.~Keysers, J.~Uszkoreit, et~al.
\newblock Mlp-mixer: An all-mlp architecture for vision.
\newblock \emph{Advances in neural information processing systems}, 34:\penalty0 24261--24272, 2021.

\bibitem[Tran et~al.(2018)Tran, Iosifidis, Kanniainen, and Gabbouj]{tran2018temporal}
D.~T. Tran, A.~Iosifidis, J.~Kanniainen, and M.~Gabbouj.
\newblock Temporal attention-augmented bilinear network for financial time-series data analysis.
\newblock \emph{IEEE transactions on neural networks and learning systems}, 30\penalty0 (5):\penalty0 1407--1418, 2018.

\bibitem[Tran et~al.(2021)Tran, Kanniainen, Gabbouj, and Iosifidis]{tran2021data}
D.~T. Tran, J.~Kanniainen, M.~Gabbouj, and A.~Iosifidis.
\newblock Data normalization for bilinear structures in high-frequency financial time-series.
\newblock In \emph{2020 25th International Conference on Pattern Recognition (ICPR)}, pages 7287--7292. IEEE, 2021.

\bibitem[Tsantekidis et~al.(2017{\natexlab{a}})Tsantekidis, Passalis, Tefas, Kanniainen, Gabbouj, and Iosifidis]{tsantekidis2017forecasting}
A.~Tsantekidis, N.~Passalis, A.~Tefas, J.~Kanniainen, M.~Gabbouj, and A.~Iosifidis.
\newblock Forecasting stock prices from the limit order book using convolutional neural networks.
\newblock In \emph{2017 IEEE 19th conference on business informatics (CBI)}, volume~1, pages 7--12. IEEE, 2017{\natexlab{a}}.

\bibitem[Tsantekidis et~al.(2017{\natexlab{b}})Tsantekidis, Passalis, Tefas, Kanniainen, Gabbouj, and Iosifidis]{tsantekidis2017using}
A.~Tsantekidis, N.~Passalis, A.~Tefas, J.~Kanniainen, M.~Gabbouj, and A.~Iosifidis.
\newblock Using deep learning to detect price change indications in financial markets.
\newblock In \emph{2017 25th European Signal Processing Conference (EUSIPCO)}, pages 2511--2515. IEEE, 2017{\natexlab{b}}.

\bibitem[Tsantekidis et~al.(2020)Tsantekidis, Passalis, Tefas, Kanniainen, Gabbouj, and Iosifidis]{tsantekidis2020using}
A.~Tsantekidis, N.~Passalis, A.~Tefas, J.~Kanniainen, M.~Gabbouj, and A.~Iosifidis.
\newblock Using deep learning for price prediction by exploiting stationary limit order book features.
\newblock \emph{Applied Soft Computing}, 93:\penalty0 106401, 2020.

\bibitem[Vaswani et~al.(2017)Vaswani, Shazeer, Parmar, Uszkoreit, Jones, Gomez, Kaiser, and Polosukhin]{vaswani2017attention}
A.~Vaswani, N.~Shazeer, N.~Parmar, J.~Uszkoreit, L.~Jones, A.~N. Gomez, {\L}.~Kaiser, and I.~Polosukhin.
\newblock Attention is all you need.
\newblock \emph{Advances in neural information processing systems}, 30, 2017.

\bibitem[Wen et~al.(2022)Wen, Zhou, Zhang, Chen, Ma, Yan, and Sun]{wen2022transformers}
Q.~Wen, T.~Zhou, C.~Zhang, W.~Chen, Z.~Ma, J.~Yan, and L.~Sun.
\newblock Transformers in time series: A survey.
\newblock \emph{arXiv preprint arXiv:2202.07125}, 2022.

\bibitem[Zeng et~al.(2023)Zeng, Chen, Zhang, and Xu]{zeng2023transformers}
A.~Zeng, M.~Chen, L.~Zhang, and Q.~Xu.
\newblock Are transformers effective for time series forecasting?
\newblock In \emph{Proceedings of the AAAI conference on artificial intelligence}, volume~37, pages 11121--11128, 2023.

\bibitem[Zhang and Zohren(2021)]{zhang2021multihorizonforecastinglimitorder}
Z.~Zhang and S.~Zohren.
\newblock Multi-horizon forecasting for limit order books: Novel deep learning approaches and hardware acceleration using intelligent processing units, 2021.
\newblock URL \url{https://arxiv.org/abs/2105.10430}.

\bibitem[Zhang et~al.(2019)Zhang, Zohren, and Roberts]{zhang2019deeplob}
Z.~Zhang, S.~Zohren, and S.~Roberts.
\newblock Deeplob: Deep convolutional neural networks for limit order books.
\newblock \emph{IEEE Transactions on Signal Processing}, 67\penalty0 (11):\penalty0 3001--3012, 2019.

\bibitem[Zhang et~al.(2020)Zhang, Zohren, and Roberts]{zhang2020deep}
Z.~Zhang, S.~Zohren, and S.~Roberts.
\newblock Deep reinforcement learning for trading.
\newblock \emph{The Journal of Financial Data Science}, 2\penalty0 (2):\penalty0 25--40, 2020.

\end{thebibliography}
\newpage
\appendix

\section{Hyperparameters Search}
\label{app:hp}
To find the best hyperparameters, we employ a grid search exploring different values as shown in Table \ref{tab:hp2}. Regarding the hyperparameters of DeepLOB and BiNCTABL, we used the ones used in \cite{prata2024lob} after a large hyperparameter search. We remark that with higher sequence sizes than 128, the performances reach a plateau. For TLOB, we also searched for the optimal number of heads, and we noted that there was not difference in between performance between 1, 2, 4, and so we fixed the number of heads to 1.
\begin{table}[!h]
\caption{The hyperparameter search spaces and best choices for each model.}
\label{tab:hp2}
\centering
\begin{tabular}{@{}llll@{}}
\toprule
Hyperparameter         & Search Space          & TLOB  & MLPLOB 
\\ \midrule
Optimizer              & \{Adam \cite{diederik2014adam}, Lion \cite{chen2024symbolic}\}  & Adam   & Adam      \\
Sequence size              & \{64, 128, 256, 384, 512\}  & 128 & 384         \\
Learning rate          & \{0.001, 0.0003, 0.0001\}  & 0.0001  & 0.003 \\
Number of layers             & \{2, 3, 4, 6\}  & 4 & 3        \\
\bottomrule
\end{tabular}
\end{table}
\newpage
\section{Additional Results}
\begin{figure*}
    \includegraphics[width=\linewidth]{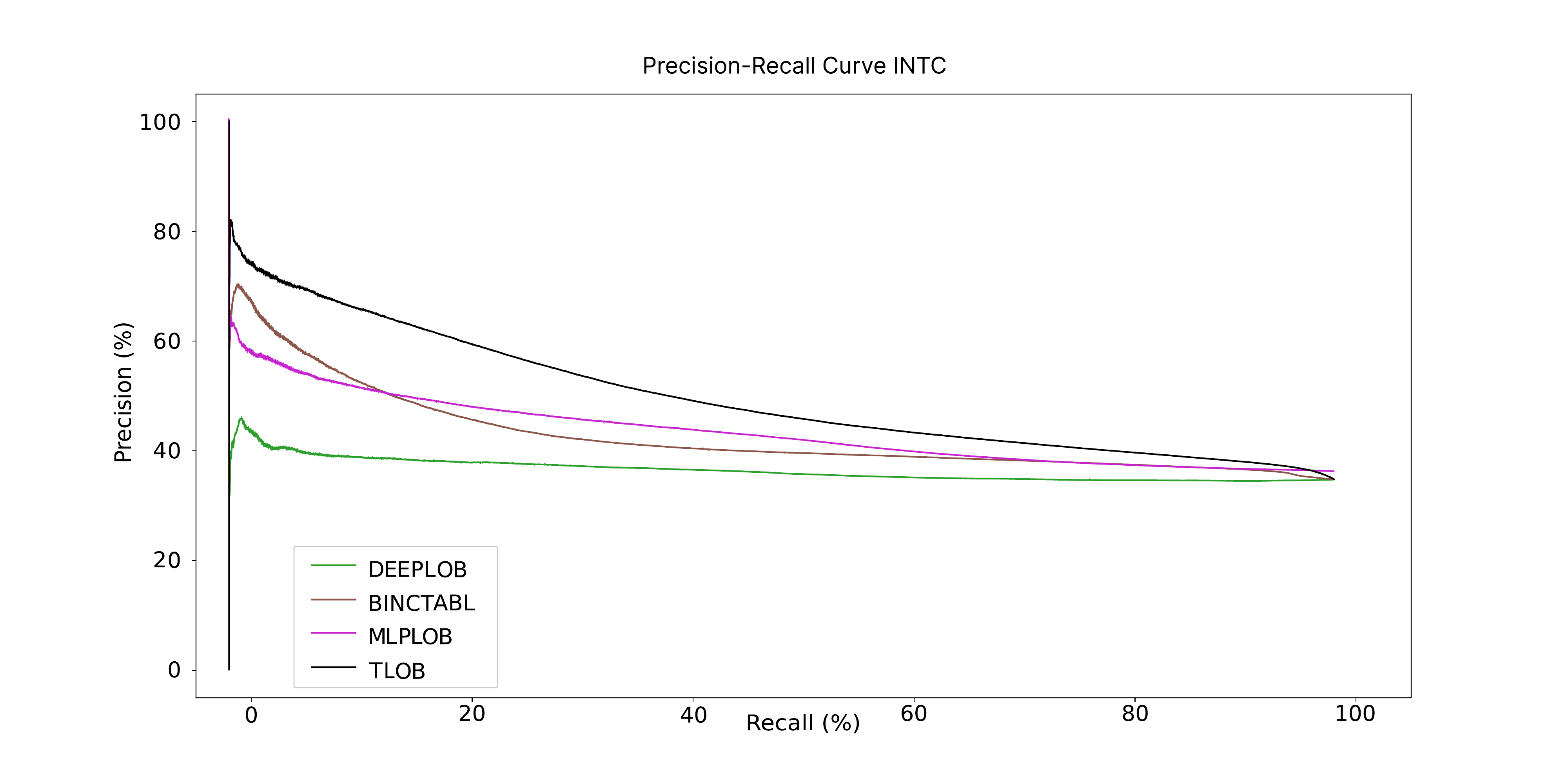}
    \caption{Precision and Recall curve for INTC for horizon = 100.}
    \label{fig:pr_intc}
\end{figure*}

\begin{figure*}[!h]
    \centering
    \includegraphics[width=\linewidth]{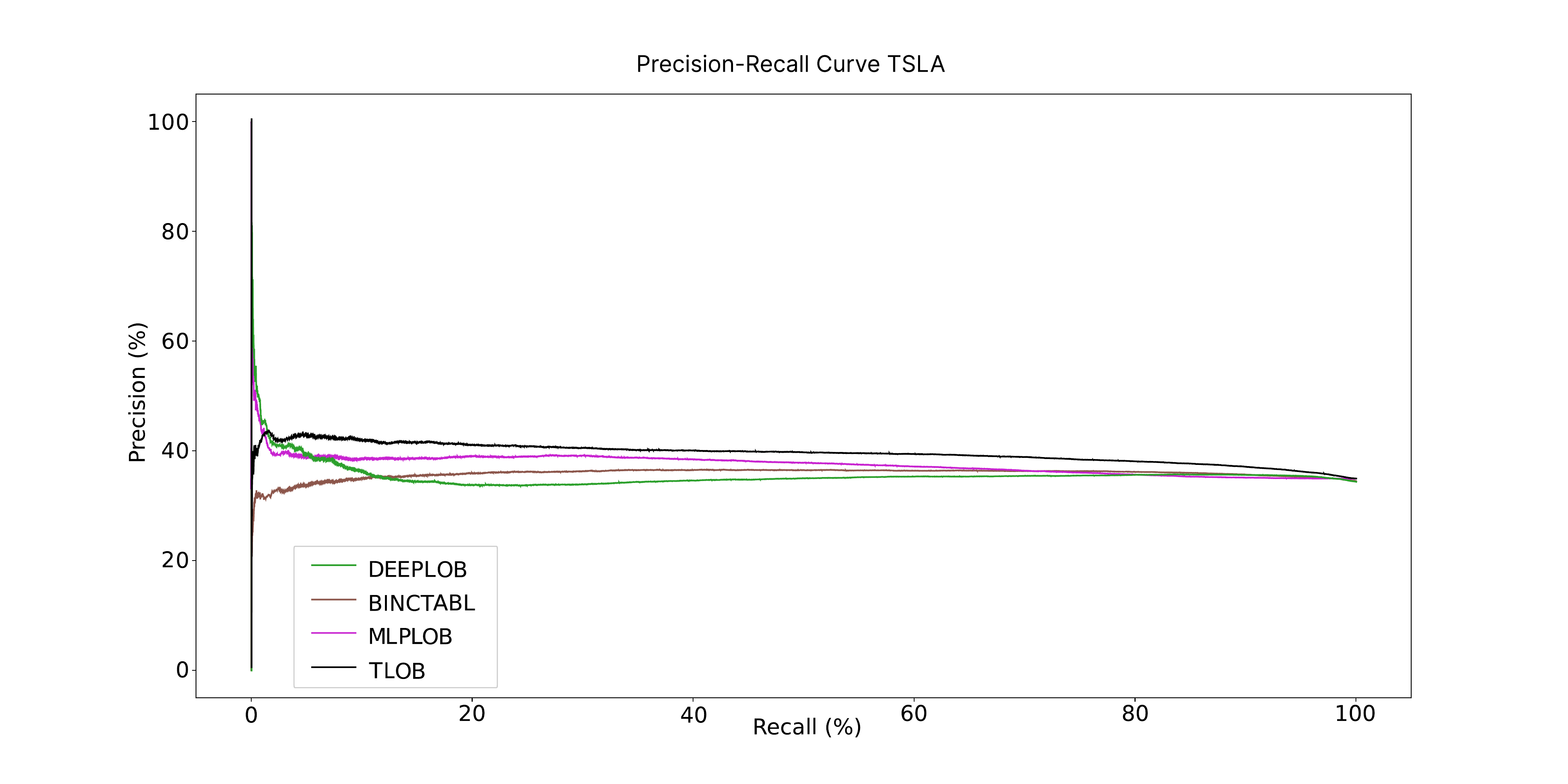}
    \caption{Precision and Recall curve for TSLA for horizon = 100.}
    \label{fig:pr_tsla}
\end{figure*}

We report the precision and recall curves for FI-2010 (Fig. \ref{fig:pr_fi}), INTC (Fig. \ref{fig:pr_intc}), and TSLA (Fig. \ref{fig:pr_tsla}), for horizon 100. As shown, across the different datasets, TLOB and MLPLOB consistently achieve higher precision at all recall levels compared to the other models. TLOB and MLPLOB, for INTC, exhibit excellent precision at low recall values, indicating their ability to accurately identify the most confident positive instances. Specifically for FI-2010, they exhibit very high precision, also at high recall values.
\label{app:results}

\end{document}